\documentclass[fleqn,twoside]{article}
\usepackage{espcrc2}
\usepackage{graphicx}

\usepackage{natbib}

 \usepackage{epsfig}
 \usepackage{subfigure}% Include figure files

\usepackage{graphicx}% Include figure files
\usepackage{epsfig}
\usepackage{dcolumn}% Align table columns on decimal point
\usepackage{bm}% bold math

\usepackage{amssymb}

\begin{document}

\title{ Study of a nonhomogeneous aerogel radiator in 
a proximity focusing RICH detector }

\author{Peter Kri\v{z}an%
\address[ULFMF]{Faculty of Mathematics and Physics, University of Ljubljana, Slovenia}%
\address[JSI]{Jo\v{z}ef Stefan Institute, Ljubljana, Slovenia}%
\thanks{Email address: peter.krizan@ijs.si.},
Samo Korpar% 
\address[UMFKKT]{Faculty of Chemistry and Chemical Engineering, University of Maribor, Slovenia}%
\addressmark[JSI],
Toru Iijima\address[NAGOYA]{Department of Physics, Nagoya University, Nagoya, Japan}}

\begin{abstract}
The use of a nonhomogeneous aerogel radiator, i.e. one consisting of layers with 
different refractive indices, 
has been shown to improve the resolution of the Cherenkov angle measured with 
a proximity focusing RICH detector. 
In order to obtain further information on the performance
of such a detector, a simple model has been used to calculate the resolution and 
search for optimal radiator parameters.
\end{abstract}

\maketitle

\section{Introduction}

As part of the upgrade of the Belle detector at KEK, it is planned to install a ring 
imaging Cherenkov detector in the forward region of the spectrometer 
to improve separation of pions and kaons in the momentum range up to 
4 GeV/c. The limited available space
has led to the decision for a proximity focusing detector using aerogel as radiator 
\cite{bib:superkekb-loi}. Different aerogel
radiators \cite{aerogel-new} as well as different position sensitive photon detectors  
\cite{flatpanel,burle} have been investigated to find an optimal solution.

An idea to further improve the resolution, has recently been proposed and experimentally
studied \cite{pk-hawaii,nim-multi}, and was later also discussed in \cite{danilyuk}. 
By  using a nonhomogeneous radiator,
i.e. multiple aerogel layers of varying refractive index, one may increase the 
number of detected photons
per per charged particle, but avoid the simultaneous increase in emission point uncertainty that 
would follow from an increase in the thickness of a
homogeneous radiator. This is achieved by suitably choosing the refractive 
indices of the consecutive layers,
so that the corresponding Cherenkov rings either overlap on the photon detector 
(focusing configuration) 
or they are well separated (defocusing configuration).
Various configurations of aerogel radiators have been experimentally 
investigated and have shown the expected 
improvement in resolution \cite{nim-multi}.

In order to achieve optimal performance of the detector in the focusing configuration, 
we have studied the influence of various radiator
parameters, such as difference in refractive index between the layers, their thickness
and transmission length, on the 
resolution of the Cherenkov angle
measured for a charged particle of given momentum. Using a simple model to 
calculate this resolution,
we have attempted to find a set of radiator parameters that would produce the 
best results, i.e. the lowest
standard deviation of the measured Cherenkov angle due to monoenergetic pions 
or kaons.

\section{The model}

 \begin{figure}[phbt]
   \begin{center}
     \epsfig{file=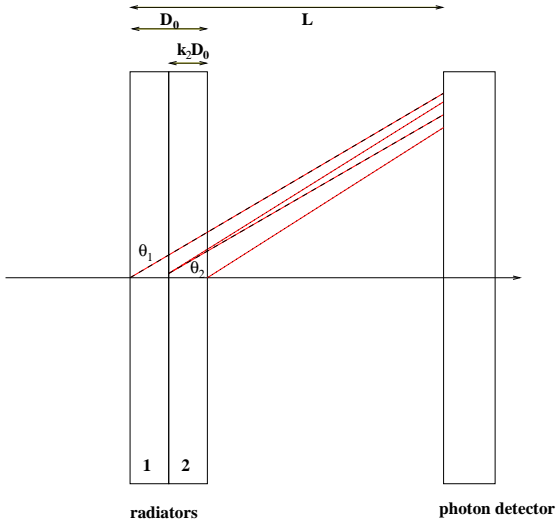,width=0.99\linewidth,clip=}
   \end{center}
   \vspace{-1cm}
   \caption{The set up of the RICH detector with a dual aerogel radiator.}
   \label{setup1}
 \end{figure}

The detector (Fig.~\ref{setup1}) has a double layer aerogel radiator of total thickness $D_0$, 
with the thickness of the downstream radiator layer (labeled 2 in Fig.~\ref{setup1}) given as $k_2 D_0$.
Refractive indices $n_1$ and $n_2=n_1+\delta n_2$ correspond to
Cherenkov angles $\Theta _1$ and $\Theta _2$.
The photon detector,  
at a distance $L$ from the entry surface of the upstream radiator, has square photosensitive pixels with the 
side equal to $\Delta$.
 \begin{figure}[phbt]
   \begin{center}
     \epsfig{file=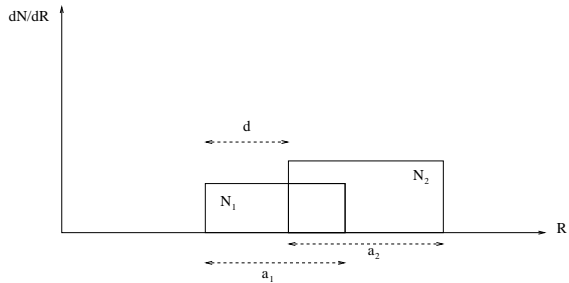,width=0.99\linewidth,clip=}
   \end{center}
   \vspace{-1cm}
   \caption{Contributions from the two radiator layers to the radial 
distribution of photon impact points on the detector plane. }
   \label{distrib}
 \end{figure} 

For a charged particle passing through the two-layer radiator, in general two 
rings are seen at the plane of the
photon detector. We assume that the distribution of 
Cherenkov photons in each ring is uniform
in the distance R from the ring center (Fig.~\ref{distrib}). This approximation is good for normal incidence and
high transmission lengths of both radiator layers.
The two uniform distributions contain $N_1$ and $N_2$ detected photons. The 
numbers of detected photons are assumed to be:
\begin{eqnarray}
N_1 &=& N_0 D_1 \sin ^2 \Theta _1 \exp(-\frac{D_1}{2 \Lambda _1}-\frac{D_2}{\Lambda _2})  \\ \nonumber
N_2 &=& N_0 D_2 \sin ^2 \Theta _2 \exp(-\frac{D_2}{2 \Lambda _2}) ,
\end{eqnarray}
where $D_2 = k_2 D_0$, $D_1 = (1-k_2)D_0$ and 
$\Lambda _1, \Lambda _2$ are the transmission lengths\footnote{In 
this model we neglect the wavelength dependence of the aerogel transmission length. 
This approximation turns out to be sufficiently good, provided the transmission length
at 400~nm is used for $\Lambda$.} of both aerogel layers. For $N_0$, the figure of merit 
of Cherenkov counters, we assume a value of 50/cm, which was a typical value from 
experimental tests of such a configuration \cite{flatpanel}.

It may be quickly seen that the rms of the distribution of photons from both 
layers (Fig.~\ref{distrib}) is given by:
\begin{eqnarray}
\sigma _R ^2 &=& \langle R^2 \rangle - \langle R \rangle ^2\\ \nonumber
&=& {1 \over 12 (N_1 + N_2)^2} \lbrace -3 \lbrack a_1 N_1 + (a_2 + 2d)N_2 
\rbrack ^2 \\ \nonumber
&+& 4(N_1 +N_2) \lbrack a_1 ^2 N_1 + (a_2 ^2 +3 a_2 d +3 d^2)N_2 \rbrack 
\rbrace ,
\end{eqnarray}
where $a_1$ and $a_2$ are the differences of outer and inner radii for the two 
rings:
\begin{eqnarray}
a_1 = D_1 \tan \Theta _1, \hspace{1cm} a_2 = D_2 \tan \Theta _2,
\end{eqnarray}
 and $d$ is the difference between inner radii of the two rings:
\begin{eqnarray}
d = (L-D_0) (\tan \Theta_{10} - \tan \Theta_{20}) + D_2 \tan \Theta_{12}.
\end{eqnarray}
Here $\Theta_{10}$ and $\Theta_{20}$ are the values of photon angles after refraction
into the air, while $\Theta_{12}$ is the angle of photons from radiator 1 in  
radiator 2.

The contribution of the emission point uncertainty to the error in 
determination 
of the Cherenkov angle due to a single photon is:
\begin{eqnarray}
\sigma _{\rm emp} ={ \sigma _R \over (L-D_0/2)}  \cos^2 \bar \Theta,
\end{eqnarray}
where we have denoted by $\bar \Theta$ the average Cherenkov angle.
The contribution due to position 
resolution of the detector, i.e. the pixel size $\Delta$, is:
\begin{eqnarray}
\sigma _{\rm det} = {\Delta \over (L-D_0/2) \sqrt{12}}  \cos^2 \bar \Theta.
\end{eqnarray}  
Other contributions, such as uncertainty of track direction or nonuniformity of density and thickness of the 
aerogel, we collectively label simply as $\sigma _{\rm rest}$. Assuming that the contributions are
not correlated, we add them in quadrature and divide by the 
square root of the number of detected photons to
obtain the rms of the Cherenkov angle for a track, i.e. the 
parameter that needs to be minimized, as
\begin{eqnarray}
\sigma _{\rm track} = {1 \over \sqrt{N_1+N_2}} \sqrt{\sigma_{\rm emp}^2 + 
\sigma_{\rm det}^2 + \sigma_{\rm rest}^2}.
\end{eqnarray}

The optimization procedure described below refers mainly to the parameters of the aerogel radiator layers
such as refractive index, thickness and transmission length. Other parameters, which are determined by
our given experimental arrangement, are mainly fixed at values corresponding to the particular
detector under study. For example, the radiator-to-photon-detector distance is 20~cm, the photon detector
pixel size is set to 6~mm and the contribution from other sources is found to be $\sigma_{\rm rest} = 6$~mrad.

\section{Results}
\subsection{Double layer radiator}

\begin{figure}[phbt]
   \begin{center}
     \epsfig{file=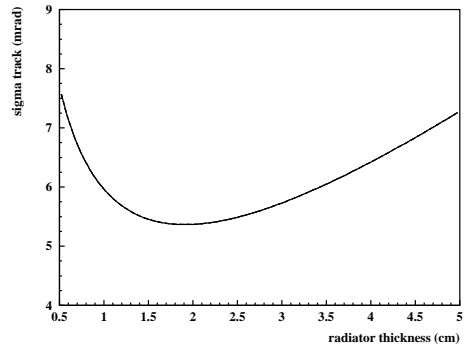,width=0.99\linewidth,clip=}
   \end{center}
   \vspace{-1cm}
   \caption{ Resolution of Cherenkov angle ($\sigma _{\rm track}$) versus the thickness $D_0$ of a homogeneous 
radiator with $n=1.04$ ($\Theta = 278$~mrad), $\Lambda=4$~cm, $L=20$~cm, $\Delta = 6$~mm and 
 $\sigma_{\rm rest} =6$~mrad.}
   \label{sigtr-single-rad-d}
 \end{figure}

First we have checked the equations and the calculation by taking  
both radiator layers of
equal refractive index ($\delta n_2 = n_2-n_1 =0$). The refractive index was set to $n=1.04$, which corresponds
to a Cherenkov angle of $\Theta = \Theta_1 = \Theta_2 = 278$~mrad for 4 GeV/c pions.
The transmission length was assumed to be $\Lambda = \Lambda _1 =\Lambda_2 = 4$~cm. The 
photon detector, at a distance of $L=20$~cm from the entry surface of
the radiator, has a fixed pad size  
$\Delta = 6$~mm. As suggested by our 
 measurements   \cite{flatpanel,pk-rich04}, the contribution from other, not yet completely understood 
sources, is $\sigma_{\rm rest}=6$~mrad.
The track resolution, 
given by $\sigma_{\rm track}$, was then calculated as a 
function of the total radiator thickness $D_0$.
The result, shown in Fig.~\ref{sigtr-single-rad-d}, gives an optimal total thickness of about 2~cm, 
at which the resolution amounts to $\sigma _{\rm track} = 5.4$~mrad. This is also consistent with 
the experimental value obtained from our measurements  \cite{flatpanel}. 

\begin{figure}[phbt]
   \begin{center}
     \epsfig{file=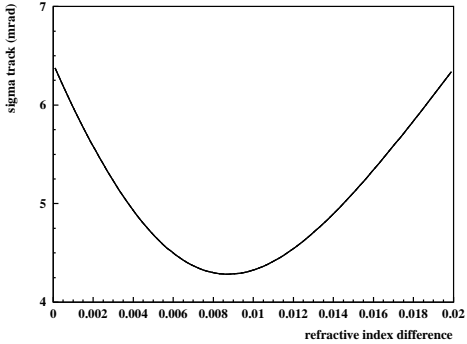,width=0.99\linewidth,clip=}
   \end{center}
   \vspace{-1cm}
   \caption{Resolution of Cherenkov angle ($\sigma _{\rm track}$) versus the difference in 
refractive indices ($\delta n_2$) of the two radiator layers. The fixed parameters are: $n_1 = 1.04$ 
($\Theta = 278$~mrad), $D_0 =4$~cm, $k_2=0.5$, $\Lambda = 4$~cm,   
$L=20$~cm, $\Delta = 6$~mm and $\sigma_{\rm rest} =6$~mrad.}
   \label{sigtr-dual-rad-dn}
 \end{figure}

In the next step, the difference $\delta n_2$ in refractive indices of the two 
radiators  has been varied for 
the case of $n_1=1.04$ ($\Theta_1 = 278$~mrad)
and two equally thick radiators ($D_0 = 4$~cm, $k_2=0.5$), with attenuation lengths 
$\Lambda _1 = \Lambda _2 = 4$~cm,
fixed distance $L = 20$~cm, standard pad size of 6~mm and $\sigma_{\rm rest} = 6$~mrad. 
The minimum of 
$\sigma _{\rm track}$ was found to be about 4.3~mrad
at a difference in refractive indices of 0.009, which corresponds to a 
difference in Cherenkov angle of $\Theta_2 - \Theta_1 = 29$~mrad %%%% ?
(Fig.~\ref{sigtr-dual-rad-dn}). We note that the minimum in $\sigma _{\rm track}$ is
quite broad, a departure of $\delta n_2$  by $\pm 0.002$ from the optimal value only
increases  $\sigma _{\rm track}$ by about 0.1~mrad.

 \begin{figure}[phbt]
   \begin{center}
     \epsfig{file=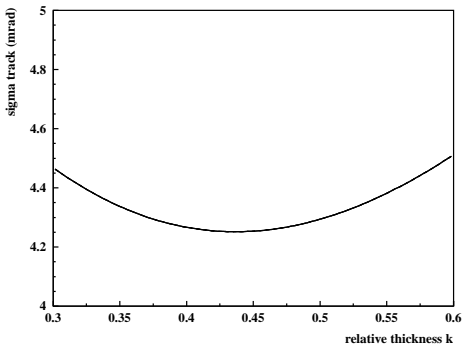,width=0.99\linewidth,clip=}
   \end{center}
   \vspace{-1cm}
   \caption{ Resolution of Cherenkov angle ($\sigma _{\rm track}$) versus relative
 thickness  $k_2 = D_2/D_0$ of the
downstream radiator layer. The fixed parameters are:  $D_0=4$~cm, $n_1 =1.04$,  
$\delta n_2=0.009$,   
$\Lambda=4$~cm, $L=20$~cm, $\Delta = 6$~mm and $\sigma_{\rm rest} =6$~mrad.}
   \label{sigtr-dual-rad-k}
 \end{figure} 

Then, $\delta n_2$ was set to 0.009 and, with all the other parameters left unchanged, 
the relative thickness of
aerogel 2 was varied. The minimum is at $k_2=0.44$ as may be seen in 
Fig.~\ref{sigtr-dual-rad-k}. 
We observe that the variation of $\sigma _{\rm track}$ with $k_2$ is weak:
it stays within 3\% of the minimum value over a broad interval $0.35 < k_2 <0.55$.

\begin{figure}[phbt]
   \begin{center}
     \epsfig{file=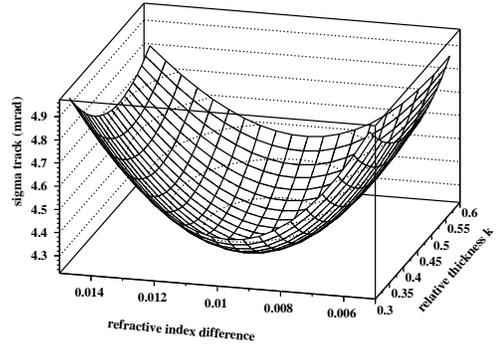,
width=0.99\linewidth,clip=}
   \end{center}
   \vspace{-1cm}
   \caption{ The resolution in Cherenkov angle ($\sigma _{\rm track}$) versus the difference in 
refractive indices $\delta n_2$ 
of the two layers and versus the relative thickness of the second layer ($k_2 = D_2/D_0$). 
The fixed parameters are: $n_1 =1.04$,
$D_0=4$~cm, $\Lambda=4$~cm, $L=20$~cm,$\Delta = 6$~mm and $\sigma_{\rm rest} =6$~mrad.}
   \label{sigtr-dual-rad-dn-k}
\end{figure}

A plot of $\sigma _{\rm track}$
depending on both relative thickness $k_2$ and difference of refractive indices 
$\delta n_2$, gives a minimum of 4.2~mrad
at $\delta n_2 = 0.009$ and $k_2=0.44$ (Fig.~\ref{sigtr-dual-rad-dn-k}). 
If, in addition to $\delta n_2$ and  $k_2$,
also the total radiator thickness $D_0$  is varied, $\sigma _{\rm track}$ has the
same minimal value (4.2~mrad) at the thickness of   $D_0=3.2$~cm.  
\begin{table}
\vspace*{1cm}
\begin{center}
\begin{tabular}{|c|c|c|c|} \hline
 & $\Lambda = \infty$ &  $\Lambda =4$~cm & $\Lambda = 3$~cm  \\ \hline
%                   & $L = 20$~cm& $L = 20$~cm & $L= 20$~cm \\ \hline
$\delta n_2$                       &0.009 & 0.007 & 0.006  \\ \hline
$D_0$ (cm)                         &  4.2 & 3.2 & 3.0  \\ \hline
$k_2$                              & 0.47 &0.45 & 0.45  \\ \hline
$\sigma_0$ (mrad)                & 14.2 & 12.5 & 12.2  \\ \hline
$N$                              & 17.4 & 9.0 & 7.7  \\ \hline
$\sigma _{\rm track} ^{\rm min}$ (mrad) & 3.4 & 4.2 & 4.4  \\ \hline
\end{tabular}
\end{center}
\caption{Optimized parameters for different $\Lambda$ and for $n_1 =1.04$, $L= 20$~cm, $\Delta = 6$~mm and 
$\sigma_{\rm rest} = 6$~mrad. The parameters are mainly explained in the text except for $N$ and  $\sigma_0$,
which are the number of photons per track and the single photon resolution, respectively.}
\end{table}
\begin{table}
\vspace*{1cm}
\begin{center}
\begin{tabular}{|c|c|c|} \hline
                   & $\Lambda = \infty $  &  $\Lambda = 4$~cm  \\ \hline
                             & $L = 25$~cm& $L = 25$~cm  \\ \hline
$\delta n_2$                        &0.008 & 0.006  \\ \hline
$D_0$ (cm)                           & 4.8 & 3.5  \\ \hline
$k_2$                               & 0.48 &0.45  \\ \hline
$\sigma_0$ (mrad)                & 12.4 & 10.8   \\ \hline
$N$                              & 19.4 & 9.4  \\ \hline
$\sigma _{\rm track} ^{\rm min}$ (mrad) & 2.8 & 3.5  \\ \hline
\end{tabular}
\end{center}
\caption {Optimized parameters for $L= 25$~cm, $n_1 =1.04$, $\Delta = 6$~mm and $\sigma_{\rm rest} = 6$~mrad.  }
\end{table}

Table 1 shows the optimized parameters for different values of the transmission length.
It is seen that aerogels with good transmission may considerably improve the resolution. 
Table 2 gives the optimized parameters for the case that 5~cm more space is available between radiator and
photon detector. The beneficial effect of these 5~cm of additional space is comparable to the effect of
perfect transmission.

We have therefore seen, that for the limited space between radiator and photon detector ($L= 20$~cm), for given
transmission length ($\Lambda= 4$~cm) and pixel size ($\Delta = 6$~mm) and 
for the given contribution of other sources to  $\sigma_{\rm track}$ ($\sigma_{\rm rest} = 6$~mrad),
one may achieve an improvement in resolution of about 1.2~mrad (i.e. from 5.4~mrad to 4.2~mrad) by optimizing
the thicknesses and refractive indices of two consecutive aerogel layers. We note that a similar 
improvement was also observed in experimental tests of such a counter \cite{nim-multi}.
Further improvements may be achieved  with aerogels of better 
transmission or with additional space available for the detector. %???
%, with smaller pixels of the photon detector or by reducing $\sigma_{\rm rest}$.

\subsection{Multiple layer radiator}

The calculation has been extended to the case when the radiator consists of more than two layers. 
Table 3 shows the optimized parameters for 3 and 4 layers compared to the dual radiator. It is evident 
that the improvement of resolution is primarily due to an increased radiator thickness, and
consequently to the 
measured number of photons per track, while the single photon resolution $\sigma_0$
remains approximately constant.

\begin{table*} 
\vspace*{1cm}
\begin{center}
\begin{tabular}{|c|c|c|c|c|} \hline
   & single layer & two layers & three layers & four  layers\\ \hline
$\delta n_2$                     &     & 0.007 &0.007 & 0.008  \\ \hline
$\delta n_3$                     &     & &0.014 & 0.015  \\ \hline
$\delta n_4$                     &         & &      &  0.022 \\ \hline
$k_2$                            &    &0.45 & 0.34 & 0.28  \\ \hline
$k_3$                            &       & & 0.27 &  0.23  \\ \hline
$k_4$                            &         &   &   &  0.19  \\ \hline
$D_0$ (cm)                       & 1.9    &  3.2 & 4.4 & 5.6  \\ \hline
$\sigma_0$ (mrad)                & 12.8   & 12.5 & 12.6 & 12.8  \\ \hline
$N$                              & 5.7   &  9.0 & 11.9 & 14.7  \\ \hline
$\sigma _{\rm track} ^{\rm min}$ (mrad)& 5.4 & 4.2  & 3.7  &    3.3  \\ \hline
\end{tabular}
\end{center}
\caption{Optimized parameters for one, two, three and four radiator layers and with fixed $n_1 =1.04$, 
$\Lambda = 4$~cm, $L= 20$~cm,  $\Delta = 6$~mm and $\sigma_{\rm rest}=6$~mrad; $k_i$ is the
relative thickness of the layer $i$, and $n_1+\delta n_i$ is the corresponding refractive index.
}
\end{table*}

 \begin{figure}[phbt]
   \begin{center}
     \epsfig{file=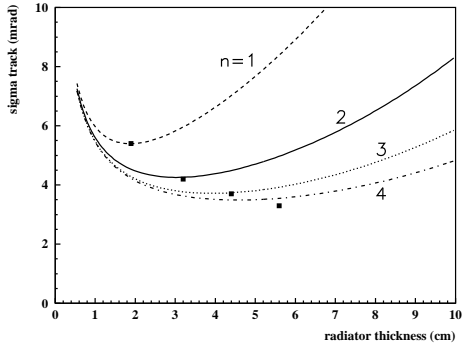,width=0.99\linewidth,clip=}
   \end{center}
   \vspace{-1cm}
   \caption{The resolution in Cherenkov angle ($\sigma_{\rm track}$) as a function of total radiator
 thickness with the number 
of layers of differing refractive index as parameter. The curves are obtained by simply dividing 
$\sigma_{\rm emp}$ with the number
of layers. The points are calculated as described in the previous section. 
The fixed parameters are: $n_1 =1.04$,
$\Lambda = 4$~cm, $L= 20$~cm, $\Delta = 6$~mm and $\sigma_{\rm rest} = 6$~mrad.}
   \label{simple-comp}
 \end{figure} 

\begin{figure}[phbt]
   \begin{center}
     \epsfig{file=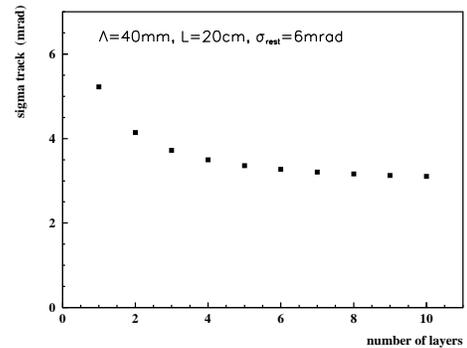,width=0.99\linewidth,clip=}
   \end{center}
   \vspace{-1cm}
   \caption{The track resolution $\sigma_{\rm track}$ at optimal $D_0$ versus the number of
 layers obtained with the 
simple estimate described in the text. The fixed parameters are: 
$n_1 =1.04$, $\Lambda = 4$~cm, $L = 20$~cm, $\Delta = 6$~mm and
 $\sigma_{\rm rest} =6$~mrad. }
   \label{simple-sigtr-nlay}
 \end{figure}

A simple estimate of $\sigma_{\rm track}$ may be obtained just by dividing the 
contribution due to emission point uncertainty
by the number of layers and adding the other (fixed) contributions in quadrature. 
The curves in Fig.~\ref{simple-comp} are obtained
by such a simplified procedure. It is seen that the results of optimization using 
the model described in section 2. of this
paper and represented by points in Fig.~\ref{simple-comp}, agree quite nicely 
with such a simple estimate. From Fig.~\ref{simple-comp} we also see that an
increased number of layers leads to an increase in optimal overall radiator thickness. 
%This dependence, obtained
%from the simple estimate is shown in Fig.~\ref{simple-d0opt} for $\Lambda = 4$~cm, $L= 20$~cm, 
%$\Delta = 6$~mm and $\sigma_{\rm rest}=6$~mrad. 
For this simplified model we plot in Fig.~\ref{simple-sigtr-nlay} the
track resolution $\sigma_{\rm track}$ at the optimized radiator thickness as a function of
the number of layers. 

We observe that the dependence of $\sigma_{\rm track}$ on radiator thickness at 
the minimum is relatively weak. It is also not
symmetric with respect to the minimum, i.e. an increase in overall thickness produces a smaller 
change in $\sigma_{\rm track}$ than an equal decrease (Fig.~\ref{sigtr-single-rad-d}
 and Fig.~\ref{simple-comp}). It would therefore be advisable to use a
 thickness, which at higher charged particle momenta is somewhat greater than the value 
required to minimize $\sigma_{\rm track}$. This would produce
a negligible loss of resolution at high momenta, but would represent a valuable increase in the 
number of detected photons at low momenta, where the difference in Cherenkov angle between 
pions and kaons is large anyway (or where 
kaons are below the threshold for Cherenkov radiation). This low 
sensitivity on the thickness of individual radiator layers (Fig.~\ref{sigtr-dual-rad-k})
also has a practical advantage; it permits equal thicknesses of the 
different layers, thus simplifying the production process.

Finally we note that the quantity we are actually interested in is the separation between pions and kaons at a given
momentum, $s_{\pi K}=(\bar \Theta_{\pi}-\bar \Theta_{K})/\sigma_{\rm track}$. Since in addition to 
 $\sigma_{\rm track}$
also the average Cherenkov angle $\bar \Theta$ varies with $k_i$, $n_i$ and $D_0$, these parameters assume
different values if $s_{\pi K}$ is optimized instead of $\sigma_{\rm track}$. The total radiator thickness is, 
e.g. somewhat 
smaller if  $s_{\pi K}$ is maximized ($D_0=3.0$~cm, 3.7~cm and 4.3~cm for the two, three and four layer configuration,
respectively). It turns out, however, that the resulting   $\sigma_{\rm track}$ and $s_{\pi K}$
 are almost equal in both cases. 

\section{Conclusion}

We have shown that with a nonhomogeneous aerogel radiator in a proximity focusing Cherenkov ring imaging
detector, one may, by adjusting the refractive index of consecutive radiator layers, reduce the contribution of
emission point uncertainty to the overall Cherenkov angle resolution of a charged particle track.
In the particular case of the RICH detector studied for the Belle upgrade, the limiting factors are then 
the lack of space and the pixel size. 
We have also seen that the optimal resolution is not very sensitive to minor variations of
refractive index or thickness of individual aerogel layers. This is a very welcome property,
since it somewhat reduces the stringent demands on the aerogel production process.

In our previous experimental studies, we have seen that there 
also exists an important contribution ($\sigma_{\rm rest}$) from factors not yet understood. These will 
be the subject of further study, as well as the investigation into 
possibilities of improving the aerogel production procedure in order to obtain the best possible transmission
and homogeneity of individual aerogel layers. However, already at this stage, the resolution in Cherenkov angle
of charged particles seems to meet the requirements of a $4 \sigma$ separation of pions and kaons
in the 1-4 GeV/c momentum range.

\end{document}